\documentclass[twocolumn,aps,prd,showpacs,nofootinbib]{revtex4}
\usepackage{graphicx}
\usepackage{amsmath}
\usepackage{amssymb}
\usepackage{bm}
\begin{document}
\title{\mbox{}\\[10pt] Line Shapes of the $\bm{Z(4430)}$}
\author{Eric Braaten}
\author{Meng Lu}
\affiliation{Physics Department, Ohio State University, Columbus, Ohio
  43210, USA}
\date{\today}
\begin{abstract}
The Belle Collaboration recently discovered the first manifestly
exotic meson: $Z^+(4430)$, which decays into $\psi'\pi^+$ and
therefore has quark content $c\bar c u \bar d$. The
proximity of its mass to the $D_1\bar{D}^*$ threshold has motivated
the interpretation of the $Z^+$ as a charm meson molecule
whose constituents are an $S$-wave superposition of
$D_1^+\bar{D}^{*0}$ and $D^{*+}\bar{D}_1^0$. If this interpretation is
correct, the small ratio of the binding energy of the $Z^+$ to the
width $\Gamma_1$ of its constituent $D_1$ can be exploited to predict
properties of its line shapes. Its full width at half maximum in the
channel $\psi'\pi^+$ should be approximately
$\sqrt{3}\Gamma_1\approx35$~MeV, which is consistent with
the measured width of the $Z^+$. The $Z^+$ should also decay into
$D^*\bar{D}^{*}\pi$ through decay of its constituent $D_1$. The peak
in the line shape for $D^*\bar{D}^*\pi$ should be at a higher energy
than the peak in the line shape for $\psi'\pi^+$ by about
$\Gamma_1/\sqrt{12}\approx 6~$MeV. The line shape in $D^*\bar{D}^*\pi$
should also be broader and asymmetric, with a shoulder on the high
energy side that can be attributed to a threshold enhancement in the
production of $D_1\bar{D}^{*}$.
\end{abstract}


\maketitle


{\it Hadrons} are bound states of quarks and antiquarks that feel
the strong force of quantum chromodynamics.
The ordinary hadrons are {\it baryons} which contain 3 quarks,
{\it antibaryons} which contain 3 antiquarks, and 
ordinary {\it mesons} which
contain a quark and an antiquark. Any other type of hadron is called
{\it exotic}.  Hundreds of hadrons have been
discovered, but none were manifestly exotic until the
recent discovery of the $Z^\pm(4430)$ by the Belle
Collaboration \cite{:2007wga}. The $Z^+$ decays into $\psi'
\pi^+$, where $\psi'$ is a charmonium meson
whose constituents are a charm quark and antiquark ($c\bar c$) 
and $\pi^+$ is a charged pion whose
constituents are a light quark and antiquark ($u\bar d$). 
Thus the $Z^+$ is a manifestly
exotic meson with quark content $c\bar c u \bar{d}$.

Since the $Z^+$ contains two quarks and two antiquarks, 
it can be called a {\it tetraquark} meson. However its structure 
is still an open question.  The
proposed interpretations of the $Z^+$ can be classified according to
the clusterings of its constituents that appear in the dominant component
of the wave function. If the $Z^+$ is a charm meson molecule
\cite{Rosner:2007mu, Meng:2007fu,
Bugg:2007vp, Ding:2007ar, Li:2007bh, Liu:2007bf}, the dominant
clusters are charm mesons with quark contents $c\bar u$ and
$\bar c d$. If the $Z^+$ is a tetraquark composed of a diquark and
an antidiquark \cite{Maiani:2007wz,Gershtein:2007vi}, the
dominant clusters are $cu$ and $\bar c\bar d$. If the $Z^+$ is a
baryonium \cite{Qiao:2007ce}, the dominant clusters are $cuq$ and
$\bar c\bar d \bar q$, where $q$ is another light quark.

The mass of the $Z^+$ is extremely close to the $D_1 \bar{D}^*$
threshold, where $D_1$ and $D^*$ are charm mesons with spin-parity 
quantum numbers $J^P = 1^+$ and $1^-$. 
This has motivated the proposal that $Z^+$ is a charm meson molecule
whose constituents are a superposition of the charm meson pairs
$D_1^+\bar{D}^{*0}$ and $D^{*+}\bar{D}_1^0$ \cite{Rosner:2007mu, Meng:2007fu,
Bugg:2007vp, Ding:2007ar, Li:2007bh, Liu:2007bf}. 
Rosner argued that the quantum numbers $J^P=0^-$ are
favored because of the small energy release in the decay
$B^+ \to K^0 Z^+$ \cite{Rosner:2007mu}. Meng and Chao used a model of
charm meson rescattering to argue that decays of $Z^\pm$ into
$D^*\bar{D}^*\pi$ should dominate over charmonium decay modes such as
$\psi' \pi^\pm$ \cite{Meng:2007fu}. Their model can accomodate a
substantial suppression of $J/\psi \, \pi^\pm$ relative to $\psi'\pi^\pm$
if $J^P=1^-$. Bugg suggested that the $Z^\pm$ should
produce a cusp at the $D_1\bar{D}^*$ threshold in several other charm
meson decay channels \cite{Bugg:2007vp}. Liu, Liu, Deng, and Zhu used
a variational approximation to argue that the one-pion-exchange
potential between $D_1$ and $\bar{D}^*$ is not strong enough to bind
them into $Z^\pm$ \cite{Liu:2007bf}. Their results are not
conclusive because their variational wavefunctions did not allow a
significant contribution to the energy from the short-distance region
where the potential is dominated by a $1/r^3$ term. 

If $Z^+$ is a $D_1 \bar D^*$ molecule, the simplest possibility 
for its orbital angular momentum state is S-wave, in which case 
its $J^P$ quantum numbers must be $0^-$, $1^-$, or $2^-$. 
The tiny binding energy of the $Z^+$ implies a strong resonant
interaction between the charm mesons that produces an
S-wave scattering length that is large
compared to the range of their interaction.
S-wave threshold resonances have universal properties that are 
determined by the large scattering length and are otherwise 
insensitive to shorter distance scales \cite{Braaten:2004rn}. 
The width $\Gamma_1$ of the $D_1$ smears out the resonant interactions
so that the universal features are not as distinctive.
However $\Gamma_1$ is small enough that we can exploit the 
universal features of the $Z^+$ provided we take into 
account the width of its constituent.

The line shape of a resonance is the invariant mass
distribution of its decay products. In this paper, we point out
simple features of the line shapes of $Z^\pm$ that follow from its
identification as a weakly-bound $S$-wave charm meson molecule. 
Its line shape in $D^*\bar{D}^*\pi$ will differ
significantly from its line shape in $\psi'\pi^\pm$ because the
resonance overlaps with a threshold enhancement in the production of
$D_1\bar{D}^*$. As a consequence, measurements of the mass and width
of the resonance in the $D^*\bar{D}^*\pi$ channel will give larger
values than in the $\psi'\pi$ channel.

A similar phenomenon has been observed in the case of the $X(3872)$,
which was discovered by the Belle Collaboration in
2003 \cite{Choi:2003ue}. Its mass is extremely close to the
$D^{*0}\bar{D}^0$ threshold. This has motivated its interpretation as
a charm meson molecule whose constituents are a superposition of
$D^{*0}\bar{D}^0$ and $D^0\bar{D}^{*0}$. In this case, it would be an
exotic meson with quark content $c\bar c u \bar u$. Unlike the
$Z^\pm$, it is not manifestly exotic because it can mix with
$c\bar c$ mesons through annihilation of the $u\bar u$ pair. The
$X(3872)$ was discovered in the decay mode $J/\psi \, \pi^+ \pi^-$. It
can also decay into $D^0\bar{D}^0\pi^0$ through the decay of its
constituent $D^{*0}$ or $\bar{D}^{*0}$. The measured mass of the 
resonance is higher in the $D^0\bar{D}^0\pi^0$ channel than in 
the $J/\psi \, \pi^+ \pi^-$ channel by about 4~MeV 
\cite{Gokhroo:2006bt,Babar:2007rva}.  The width of the $X$ resonance
also appears to be larger in the $D^0\bar{D}^0\pi^0$ channel.  
The width in the
$D^0\bar{D}^0\pi^0$ channel was measured to be $2.42 \pm 0.55$~MeV
\cite{Gokhroo:2006bt}, but there is only an upper bound on
the width in the $J/\psi \, \pi^+\pi^-$ channel: $\Gamma_X < 2.3$~MeV
at 90\% confidence level \cite{Choi:2003ue}. In
Refs.~\cite{Hanhart:2007yq,Braaten:2007dw}, the line shapes of
$X(3872)$ were analyzed under the assumption that it is a charm meson
molecule. The observed mass difference was shown to be compatible 
with this interpretation. An alternative
explanation of the mass difference between the two channels is that
they correspond to two nearly degenerate tetraquark mesons
\cite{Maiani:2007vr}.

We begin by summarizing the properties of the $Z^+(4430)$. It was
discovered in the decay $B^+ \to K^0 + Z^+$. 
Its measured mass and width are \cite{:2007wga}
\begin{subequations}
\begin{eqnarray}
M_Z &=& 4433 \pm 4 \pm 2~\mathrm{MeV},
\label{Eq:MZ}
\\
\Gamma_{Z} &=& 45^{+18}_{-13}{}^{+30}_{-13}~\mathrm{MeV}.
\label{Eq:GammaZ}
\end{eqnarray}
\end{subequations}
The mass in Eq.~(\ref{Eq:MZ}) is close to the
threshold for the charm mesons $D^{*+}$ and $\bar{D}_1^0$, which have
quark content $c\bar d$ and $\bar c u$ and spin-parity quantum
numbers $J^P = 1^-$ and $1^+$. The difference between charm meson
masses are measured more accurately than the masses themselves. If we
combine the PDG values of the mass differences \cite{Yao:2006px} with
a recent precise measurement of the $D^0$ mass by the CLEO
Collaboration \cite{Cawlfield:2007dw}, we find that the
$D^{*+}\bar{D}_1^0$ threshold is at
\begin{equation}
\label{Eq:D1Ds}
M_{D_1^0} + M_{D^{*+}} = 
4432.2 \pm 0.9~{\rm MeV} .
\end{equation}
This is not a sharp threshold because the $D_1^0$ has a width
$\Gamma_1 =20.4\pm 1.7~$MeV \cite{Yao:2006px}. The $D_1^+\bar{D}^{*0}$
threshold may differ by a few MeV, but the difference is negligible
compared to $\Gamma_1$. The binding energy of the $Z^+$ defined by the
difference between the threshold in Eq.~(\ref{Eq:D1Ds}) and its
measured mass in Eq.~(\ref{Eq:MZ}) is 
\begin{equation}
\label{Eq:EZ}
 E_Z = -1 \pm 5~\mathrm{MeV}.
\end{equation}
The central value is negative corresponding to a resonance or a
virtual state, but the error bar is compatible with a bound
state.  The discovery decay mode is $Z^+ \to \psi' \pi^+$.
The decay products $\psi'$ and $\pi^+$ have isospin
and G-parity quantum numbers $I^G=0^-$ and $1^-$, respectively.
Thus the quantum numbers of the $Z^+$ are $I^G=1^+$.
If the $Z^+$ is a $D_1\bar{D}^*$ molecule, 
the members of its isospin multiplet have the particle content
\begin{eqnarray}
\label{Eq:ZContent}
Z^+
&=& 
\tfrac{1}{\sqrt{2}}
\left( D^{*+} \bar{D}_1^0 - D_1^+ \bar{D}^{*0} \right),
\nonumber \\
Z^0
&=& 
\tfrac{1}{2}
\left( D_1^0 \bar{D}^{*0} - D^{*0} \bar{D}_1^0 
	+ D_1^+ D^{*-} - D^{*+} D_1^- \right),
\nonumber \\
Z^-
&=& 
\tfrac{1}{\sqrt{2}}
\left( D_1^0 D^{*-} - D^{*0}  D_1^- \right).
\nonumber 
\end{eqnarray}

If the $Z$ has $J^P$ quantum numbers $0^-$, $1^-$, or $2^-$,
it has an S-wave coupling to $D_1\bar{D}^*$.
The tiny binding energy in Eq.~(\ref{Eq:EZ}) then implies that it
is an S-wave threshold resonance with universal properties.
We proceed to describe the consequences  for elastic scattering 
of its constituents.  
The amplitude $\mathcal{A}$ for elastic scattering of a 
$D_1\bar{D}^*$ pair in a resonant S-wave channel can be written as
\begin{equation}
\mathcal{A} = (2\pi/M_{1*}) f(E)~,
\end{equation}
where $M_{1*}$ is the reduced mass for the $D_1\bar{D}^*$ pair.
The scattering amplitude $f(E)$ depends only on the energy difference
$E$ between the invariant mass of the charm mesons and the
$D_1\bar{D}^*$ threshold. If there is a threshold S-wave resonance,
the scattering amplitude for energy $E$ near the threshold
has a simple universal form:
\begin{equation}
f(E) = [- \gamma + \kappa(E)]^{-1}~,
\label{Eq:ScatteringAmplitude}
\end{equation}
where $\kappa(E) = (-2M_{1*} E - i \epsilon)^{1/2}$ and
$\gamma$ is the inverse scattering length. 
The universal expression in Eq.~(\ref{Eq:ScatteringAmplitude})
is valid when $\kappa$ is large compared to 
$\frac12 | r_s \kappa^2 |$, where $r_s$ is the S-wave effective range.
Since the long-range interactions of $D_1$ and $\bar{D}^*$ 
are dominated by pion exchange, we expect $|r_s| \lesssim 1/m_\pi$.
Thus the universal region $|E| < 2/(M_{*1} r_s^2)$ should 
at least include the interval $|E| < 36$~MeV.
The simple expression for $\kappa(E)$ with a branch point at $E=0$
applies if the constituents are stable.  
If the width $\Gamma_1$ of $D_1$ is taken into account, 
the branch point should be at $E=-i\Gamma_1/2$. 
Thus an expression for $\kappa(E)$ that takes into account 
the width of the constituent $D_1$ is \cite{Braaten:2007dw}
\begin{equation}
\label{Eq:kappaG1}
\kappa(E) = \sqrt{-2 M_{1*} \left(E + i \Gamma_1/2 \right)}.
\end{equation}
If $E$ is real, a more explicit expression for $\kappa(E)$ with
the appropriate choice of branch cut is
\begin{eqnarray}
\label{Eq:kappa-Re-Im}
\kappa(E) 
&=&   \sqrt{M_{1*}}
\left[ (E^2+\Gamma_1^2/4)^{1/2} - E\right]^{1/2} 
\nonumber \\
& & -i\sqrt{M_{1*}}
\left[ (E^2+\Gamma_1^2/4)^{1/2} + E\right]^{1/2}~.
\end{eqnarray}

The universal expression for the scattering amplitude given by
Eqs.~(\ref{Eq:ScatteringAmplitude}) and (\ref{Eq:kappa-Re-Im}) 
can be motivated by unitarity. The
imaginary part of $f(E)$ can be written as
\begin{equation}
\label{Eq:OpticalTheorem}
\textrm{Im} f(E) 
= 
\left|f(E)\right|^2 
\textrm{Im} \left[ \gamma - \kappa(E) \right]~.
\end{equation}
If $\gamma$ is real and if $\Gamma_1=0$,
$\textrm{Im}[\gamma-\kappa(E)]$ reduces to
$(2M_{1*}E)^{1/2}\theta(E)$. In this case,
Eq.~(\ref{Eq:OpticalTheorem}) is simply the optical theorem 
that expresses the exact unitarity of elastic scattering 
in the $S$-wave channel.
The effects of inelastic scattering can be taken into
account through the variables $\mathrm{Im}\gamma$ and
$\Gamma_1$. Since the dominant decay modes of the
$D_1$ are $D^*\pi$ \cite{Godfrey:2005ww}, 
the variable $\Gamma_1$ takes
into account inelastic scattering channels of the form
$D^*\bar{D}^*\pi$. The variable $\mathrm{Im}\gamma$ takes into account
other inelastic scattering channels that do not involve the decay
of a constituent of $Z$, such as $\psi'\pi$, $J/\psi \, \pi$,
$D^*\bar{D}$ and $D^*\bar{D}^*$. We refer to them as
\textit{short-distance channels}, because they proceed through
intermediate states in which the $D_1$ and $\bar{D}^*$ have a
separation that is small compared to $1/|\gamma|$. The imaginary part
of $\gamma$ can be partitioned into contributions from the individual
short-distance channels: 
\begin{equation}
\label{Eq:gamma-GammaC}
\textrm{Im}\gamma =  \textstyle{\sum_C} \, \Gamma^C .
\end{equation}

To derive universal expressions for the line shapes of the $Z^+$
in the decay $B^+ \to K^0 +Z^+$, we start from the optical theorem 
for the inclusive decay into $K^0$ plus particles that couple 
to the $Z^+$ resonance.  If the difference $E$ between the 
invariant mass of the resonating particles and the $D_1\bar{D}^{*0}$ 
threshold is small enough, the invariant mass distribution 
can be written as
\begin{equation}
\label{Eq:LineShapeResonant}
\frac{d\Gamma}{dE}\left[B^+\to K^0 + \text{resonant}\right] 
= 2 \Gamma_B^K ~ \textrm{Im}f(E)~,
\end{equation}
where $\Gamma_B^K$ is a constant with dimension of energy that
takes into account the probability for the constituents of $Z^+$ to
be created in the transition $B^+ \to K^0$. The energy
distribution in Eq.~(\ref{Eq:LineShapeResonant}) can be decomposed
into the line shapes for individual resonant channels by inserting
Eqs.~(\ref{Eq:OpticalTheorem}) and (\ref{Eq:gamma-GammaC}).
The line shape in the short-distance decay mode $\psi'\pi^+$ is
\begin{equation}
\frac{d\Gamma}{dE}\left[ B^+ \to K^0 + \psi' \, \pi^+ \right] 
= 
2\Gamma_B^K ~ \left| f(E) \right|^2 \Gamma^{\psi'\pi}~.
\label{Eq:LineShapeShort}
\end{equation}
This line shape applies equally well to any other short-distance
decay mode $C$, with the constant $\Gamma^{\psi'\pi}$ replaced by 
$\Gamma^C$. The line shape of $Z^+$ in the $D^*\bar{D}^*\pi$ channels
can be obtained by inserting Eq.~(\ref{Eq:OpticalTheorem})
into Eq.~(\ref{Eq:LineShapeResonant}) and using the expression for
$\textrm{Im}\kappa(E)$ in Eq.~(\ref{Eq:kappa-Re-Im}):
\begin{eqnarray}
\frac{d\Gamma}{dE}\left[ B^+ \to K^0 +  D^* \bar{D}^* \pi \right] 
\hspace{3.3cm}
\nonumber
\\
=
2\Gamma_B^K \left| f(E) \right|^2
\sqrt{M_{1*}}
\left[(E^2+\Gamma_1^2/4)^{1/2} + E \right]^{1/2},
\label{Eq:LineShapeLong}
\end{eqnarray}
where $D^* \bar{D}^* \pi$ denotes $D^{*+} D^{*-} \pi^+$ or
$D^{*+} \bar{D}^{*0} \pi^0$ or $D^{*0} \bar{D}^{*0} \pi^+$.  Since
$\bar{D}_1^0$ decays primarily into
$D^{*-}\pi^+$ and $\bar{D}^{*0} \pi^0$ with branching fractions
$\tfrac{2}{3}$ and $\tfrac{1}{3}$ and $D_1^+$
decays primarily into $D^{*+}\pi^0$ and $D^{*0} \pi^+$ with branching
fractions $\tfrac{1}{3}$ and $\tfrac{2}{3}$ \cite{Godfrey:2005ww}, the
line shapes in any of the three individual $D^*\bar{D}^*\pi$ channels
can be obtained by multiplying Eq.~(\ref{Eq:LineShapeLong}) by $1/3$.
The line shapes given in Eqs.~(\ref{Eq:LineShapeShort}) and
(\ref{Eq:LineShapeLong}) require only that the $Z$ has 
$J^P=0^-$, $1^-$, or $2^-$. 
The spin $J$ determines the correlations between the
polarizations of the two spin-1 constituents of $Z^+$.

We proceed to illustrate the line shapes of the $Z$.
\begin{figure}[t]
\includegraphics[width=0.45\textwidth,clip=true]{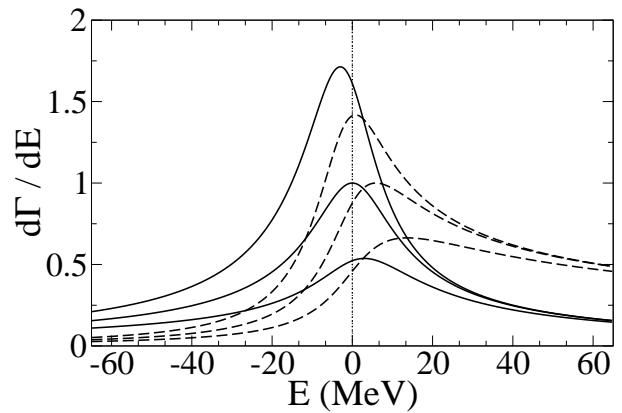}
\caption{Line shapes of the $Z^+$ in the $\psi'\pi^+$ decay
channel (solid lines) and in $D^*\bar{D}^*\pi$ decay channels (dashed
lines) for three values of the binding energy:
$E_Z = +3$~MeV (upper lines), 0~MeV (middle lines), 
and $-3$~MeV (lower lines).  The line shapes are normalized 
so that their maximum values are 1 for $E_Z = 0$.
\label{Fig:LineShapes}}
\end{figure}
We will assume that decays of $Z$ into $D^* \bar D^* \pi$
dominate over the short-distance decays.
We can therefore neglect $\mathrm{Im}\gamma$ in the resonance factor 
$|f(E)|^2$ in the line shapes for $\psi'\pi$ and $D^* \bar{D}^* \pi$
in Eqs.~(\ref{Eq:LineShapeShort}) and
(\ref{Eq:LineShapeLong}).  Thus the only unknown
parameters are $\textrm{Re}\gamma$ and the normalization factors
$\Gamma_B^K$ and $\Gamma^{\psi'\pi}$. We identify the binding energy
$E_Z$ with the negative of the position of the peak in the line shape
for $\psi'\pi$. It satisfies the equation
\begin{equation}
\label{Eq:SDEMax}
E_Z =
\frac{\textrm{Re} \gamma}{2 \sqrt{M_{1*}}}
\left[ (E_Z^2+\Gamma_1^2/4)^{1/2} + E_Z \right]^{1/2}.
\end{equation}
The line shapes of $Z^+$ in $\psi'\pi^+$ and $D^*\bar{D}^*\pi$
are shown in Fig.~\ref{Fig:LineShapes} for
three values of the binding energy: $E_Z=+3$, $0$ and $-3$~MeV, which
correspond to $\textrm{Re}\gamma=+53.9$, $0$, and $-72.0$~MeV,
respectively. The normalization factors $\Gamma_B^K$ and
$\Gamma^{\psi'\pi}$ were chosen so that the two line shapes have the
same peak value for $E_Z=0$. These same factors were then used for
$E_Z=\pm 3$~MeV. The peaks in the line shape are most dramatic if
$E_Z$ is positive. The line shapes in $\psi'\pi$ are nearly
symmetric about the peak at the energy $-E_Z$. If $E_Z < 0$, the
line shape is sometimes referred to as a ``cusp'' \cite{Bugg:2007vp}
because it has a discontinuity in slope at $E=0$ if $\Gamma_1 = 0$. 
In the case of $Z^+$, the cusp is completely smoothed out by 
the width of the $D_1$, as is evident in the lowest solid line in
Fig.~\ref{Fig:LineShapes}.
The line shapes in $D^*\bar{D}^*\pi$ have a peak at a higher energy
than for $\psi'\pi$
and they also have a broad shoulder on the high energy side of the
peak. This behavior can be attributed to a threshold enhancement
in the production of $D_1\bar{D}^*$ and $D^*\bar{D}_1$ that
overlaps with the resonance.

To quantify the behavior of the line shapes
in Eqs.~(\ref{Eq:LineShapeShort}) and (\ref{Eq:LineShapeLong}), we
can exploit the fact that the binding energy $E_Z$ is small compared
to the width $\Gamma_1$ of the constituent. Eq.~(\ref{Eq:SDEMax}) for
$E_Z$ can be solved as an expansion in powers of
$\mathrm{Re}\gamma$ whose leading term is
$[\Gamma_1/(8 M_{1*})]^{1/2} \textrm{Re} \gamma$.
The expansion can be inverted to give $\mathrm{Re}\gamma$ as an
expansion in $E_Z/\Gamma_1$. 
To first order in $E_Z/\Gamma_1$, the energy at which
the line shape for $D^*\bar{D}^*\pi$ has its maximum is
\begin{equation}
\label{Eq:LDEMaxExpan}
E_{\textrm{max}}^{D^*\bar{D}^*\pi} 
\approx 0.289 \, \Gamma_1 - 2.03 \, E_Z ~.
\end{equation}
Since $|E_Z| \ll \Gamma_1$, the difference
between the locations of the peaks in the $D^*\bar{D}^*\pi$ and 
$\psi'\pi$ line shapes is approximately 
$\Gamma_1/\sqrt{12}\approx6~$ MeV.
The line shapes in Eqs.~(\ref{Eq:LineShapeShort}) and
(\ref{Eq:LineShapeLong}) are not Breit-Wigner resonances, so they
can not be characterized simply by a width $\Gamma_Z$. 
A simple way to quantify their widths is to give the energies at
which the line shapes decrease to half of their maxima. For the
$\psi'\pi$ line shape, the half-maxima are at the energies
\begin{subequations}
\begin{eqnarray}
 E_{+}^{\psi'\pi} &\approx& 
0.866 \, \Gamma_1 - 3.42 \, E_Z~,
\\
 E_{-}^{\psi'\pi} &\approx& 
- 0.866 \, \Gamma_1 + 0.16 \, E_Z~.
\end{eqnarray}
\end{subequations}
The full width at half maximum is approximately 
$\sqrt{3}~\Gamma_1\approx 35$~MeV, 
which is consistent with the measured width of $Z^+$
in Eq.~(\ref{Eq:GammaZ}).~~ For the $D^*\bar{D}^*\pi$ line shape,
the half-maxima are at the energies
\begin{subequations}
\begin{eqnarray}
 E_+^{D^*\bar{D}^*\pi} 
&\approx&  
3.017 \, \Gamma_1 - 15.37 \, E_Z~,
\\
E_-^{D^*\bar{D}^*\pi} 
&\approx&
-0.371 \, \Gamma_1 - 1.08 \, E_Z~.
\end{eqnarray}
\end{subequations}
The full width at half-maximum is approximately
$3.4 \,\Gamma_1\approx 69$ MeV, which is about twice the width
in $\psi'\pi$. 
The right half-maximum is about 4 times 
as far from the peak energy as the left half-maximum.

The universal scattering amplitude in 
Eq.~(\ref{Eq:ScatteringAmplitude}) also applies to the $X(3872)$ 
if it is a $D^* \bar D$ molecule \cite{Braaten:2007dw}.
The difference between the
lifetimes of the constituents of the $X$ and $Z^\pm$
leads to a significant difference in their line shapes. 
In the case of the $X$, the $D^{*0}$ has a width of about 65~keV, 
which is small compared to the 8~MeV splitting   
between the $D^{*0}\bar{D}^0$ and $D^{*+}D^-$
thresholds. Since the binding energy of $X$ relative to the
$D^{*0}\bar{D}^0$ threshold is small compared to 8~MeV, 
it is a $D^{*0}\bar{D}^0$ molecule with a very small $D^{*+}D^-$ component.
The line shapes of the $X$ are given by simple universal formulas
analogous to those in Eqs.~(\ref{Eq:LineShapeShort})  
and (\ref{Eq:LineShapeLong}) only if the energy is
within a few MeV of the $D^{*0}\bar{D}^0$ threshold. 
For energies more than a few MeV from the threshold, 
one must take into account the resonant
coupling to the charged charm meson channel $D^{*+}D^-$
\cite{Voloshin:2007hh,Braaten:2007ft}.
The width of the $D^*$ provides less smearing than that of the $D_1$, 
so it is possible for the line shape for the $X$ to have a 
two-peaked structure consisting of a resonance 
and a threshold enhancement.

In summary, the hypothesis that the $Z^\pm$ is a weakly-bound
$S$-wave charm meson molecule implies that its line shape in
$D^*\bar{D}^*\pi$ should peak at a higher energy and have a larger
width than its line shape in $\psi'\pi$. If our quantitative
predictions for these line shapes are confirmed, it would provide
strong support for this interpretation of $Z$. The tiny binding energy
of the $Z$ requires an accidental fine-tuning of the charm quark
mass. The existence of the $Z$ suggests that there may be $c\bar c$
tetraquark mesons in other $J^P$ and $I^G$ channels that are more
strongly bound. It also implies that there must be a $b\bar b$
analog of the $Z$ with a much larger binding energy 
\cite{Cheung:2007wf,Lee:2007gs}. Thus the discovery of the $Z(4430)$ 
is just the beginning of the spectroscopy of tetraquark mesons.

This research was supported in part by the Department of Energy
under grant DE-FG02-91-ER40690.

\end{document}